%% file: cluster_assembly.tex
\documentclass[aip, superscriptaddress, reprint, onecolumn]{revtex4-1} %
\usepackage{natbib}
\usepackage{wrapfig} %
\usepackage{color}
\usepackage{hyperref}
\usepackage{amsmath, amssymb}
\usepackage{graphicx} %

\usepackage[margin=1in]{geometry}

\begin{document}

\title{
The emergence of bulk structure in clusters via isotropic multi-well pair potentials
}

\author{Jennifer E.\ Doyle}
\affiliation{Department of Physics and Astronomy, Wellesley College, Wellesley, MA}
\affiliation{These authors contributed equally to this work}

\author{Maya M.\ Martirossyan}
\affiliation{Department of Materials Science and Engineering, Cornell University, Ithaca, NY}
\affiliation{These authors contributed equally to this work}
\affiliation{Present address: Center for Soft Matter Research, Department of Physics, New York University, New York, NY}

\author{Julia Dshemuchadse}
\affiliation{Department of Materials Science and Engineering, Cornell University, Ithaca, NY}
\affiliation{To whom correspondence should be addressed: jd732@cornell.edu}

\author{Erin G.\ Teich}
\affiliation{Department of Physics and Astronomy, Wellesley College, Wellesley, MA}
\affiliation{To whom correspondence should be addressed: et106@wellesley.edu}

\begin{abstract} 

The mechanical, optical, and chemical properties of a wide variety of soft materials are enabled and constrained by their bulk structure.
How this structure emerges at small system sizes during self-assembly has been the subject of decades of research, with the aim of designing and controlling material functionality.
Despite these efforts, it is still not fundamentally understood how nontrivial interparticle interactions in a finite $N$-body system influence resultant structure, and how that structure depends on $N$. 
In this study, we investigate the emergence of non-close packings using multi-well isotropic pair potentials to simulate finite cluster formation of four distinct two-dimensional crystal structures. 
These pair potentials encode multiple preferred length scales into the system, allowing us to understand how anisotropic structural motifs---as opposed to close-packing---emerge as cluster size $N$ increases. 
We find a tendency toward close-packing at small system sizes irrespective of the bulk structure; however, the system size at which bulk structure emerges is influenced by the coordination number of the bulk and the shape of the pair potential. 
Anisotropic structure emerges through the formation of bonds at a secondary bonding length at larger system sizes, and it is also dependent upon the shape of the pair potential.
Our findings demonstrate that tuning particle--particle interactions can enable the engineering of nano- or mesoscale soft matter clusters, in applications as diverse as drug delivery and hierarchical materials design. 

\end{abstract}

\maketitle

\section{Introduction}

Materials functionality is highly dependent on materials structure.
This is the case not only within atomic systems, whose magnetic, electrical, mechanical, and optical properties are often entirely determined by atomic structure, but also for soft-matter systems whose constituent particles are of nano- or micrometer size.\cite{Boles2016,Deng2020}
For example, semiconductor nanocrystal superlattices have bandgaps that are dependent on particle size and configuration, with uses in light emitting devices and fluorescence probes.\cite{Talapin2010}
Nanoparticle assemblies also have plasmonic responses under incident electromagnetic fields that are heavily influenced by particle shape \cite{Mulvihill2010} and arrangement, leading to potential applications as diverse as prion \cite{Alvarez-Puebla2011} or chemical detection.\cite{Nie1997}
Magnetic nanocrystal assemblies of multiple species have unique magnetic alignment properties \cite{Chen2010a} which may be useful for data storage,\cite{Sun2000} while nanocrystals featuring catalytic particles have properties that are highly dependent on particle geometry and assembly,\cite{Kang2013} with applications in the oxidation of alcohol \cite{Auyeung2015} and methanol.\cite{Sun2015}
In each of the previous contexts, materials properties are determined by bulk crystal structure in the thermodynamic limit of extremely large system size.
In order to control bulk structure and thus design materials properties, it is crucial that we understand how that structure emerges at smaller system sizes.

Finite clusters provide an especially useful lens through which to investigate the emergence of bulk structure.\cite{Pretti2019}
The evolution of cluster structure as system size increases has been studied experimentally in diverse systems consisting of noble gas atoms,\cite{Echt1981,Harris1984} colloidal (hard-sphere) particles,\cite{Velev2000,Manoharan2003,DeNijs2014,Wang2018a} gold nanoparticles,\cite{Lacava2012} droplets,\cite{Guzowski2015} colloids with attractive interactions induced by electrostatic charge and DNA complementarity,\cite{Schade2013, Song2018} and polyhedral plasmonic nanoparticles.\cite{Henzie2013}
Cluster structure in various model systems has also been investigated theoretically for systems whose particles interact via Lennard-Jones,\cite{Wales1997} Morse,\cite{Doye1995} and Dzugutov \cite{Doye2001} potentials, hard-sphere systems,\cite{Lauga2004} sticky-sphere systems with extremely short-range interactions,\cite{Kallus2017} systems of anisotropic polyhedral particles,\cite{Teich2016,Skye2022} and systems of attractive cone-like particles.\cite{Chen2007}
In addition to their use in understanding the emergence of bulk structure, finite clusters also have applications as drug delivery vehicles,\cite{Dinsmore2002} information storage units,\cite{Phillips2014a} and building blocks for the creation of hierarchical materials \cite{Lakes1993} whose optical properties are dependent on cluster geometry,\cite{Henzie2013} with potential uses in cloaking,\cite{Pendry2006} metafluid engineering,\cite{Hinamoto2020} and photonic bandgap tuning.\cite{He2020}

In every system mentioned above, finite cluster structure---and its growth sequence on the way to the bulk---is highly dependent on the governing interactions between particles and not known \textit{a priori}.
To investigate the relationship between pairwise particle interaction and growth sequences of finite clusters, it is necessary that the sequences be examined incrementally as cluster size increases and particle pair interactions are tuned. 
Previous work \cite{Doye1995} has focused primarily on simple, single-well pair potentials with one preferred interparticle bonding distance, while other studies of finite cluster structures \cite{Wales1997, Doye2001} focus on ``magic sizes'' and ground state motifs rather than drawing conclusions about growth sequences more generally.
Here, we employ tunable particle interactions to examine how they influence growth and structure development in exhaustive and incremental finite cluster sequences in several systems.
We use a pair potential consisting of two energy wells---whose depth and distance relative to each other can be continuously changed (the so-called Lennard-Jones--Gauss (LJG) potential \cite{Engel2007a})---in order to alter pairwise bonding preferences on competing length scales.
This family of potentials was previously shown to stabilize a wide diversity of bulk structures in both two \cite{Engel2007a,Engel2008a} and three dimensions,\cite{Engel2008a,Dshemuchadse2021a} as well as finite clusters in three dimensions.\cite{Wesnak2023}
We study finite cluster growth with this interaction potential and only examine assembly in two dimensions to avoid polytetrahedral frustration \cite{Nelson1989} in three dimensions, where the close-packing of particles gives rise to tetrahedral bonding motifs that cannot periodically extend to form crystal structures.
In two dimensions, by contrast, the close-packing of particles gives rise to triangular bonding motifs which can periodically extend to form the triangular lattice.

In this paper, we use the LJG pair potential to examine the emergence of bulk structure via finite clusters in four two-dimensional systems, which were previously shown to self-assemble into the triangular, square, honeycomb, and rhombic lattices in the bulk.\cite{Engel2007a}
We also generate minimal-energy clusters for the single-well two-dimensional Lennard-Jones (LJ) system as a comparative reference for our results.
At very small system sizes, we find that all systems exhibit bonding at a single length scale and form triangular close-packed clusters similar to those found in the LJ system.
Each system, however, subsequently deviates from this close-packing at larger system sizes and instead exhibits bonding at multiple length scales in a manner that depends on the shape of the pair potential.
We find that the system sizes at which these transitions occur are also highly dependent on the bulk structure itself, with the more open bulk structures generally emerging at larger system sizes.

Collectively, our results show that the emergence of structure in finite clusters depends on both the competition induced by multiple preferred interparticle distances due to the governing interaction potential and also the bulk crystal structure ultimately self-assembled by that potential.
Tuning interaction potentials thus represents a means of designing structure in both the bulk thermodynamic limit and within finite clusters far from the thermodynamic limit.

\section{Methods}

\subsection{Simulations}
We simulated the formation of finite clusters ranging from $N=2$ to $100$ particles via molecular dynamics for four systems, each of which is governed by a different parameterization of the LJG pair potential.\cite{Engel2007a}
The LJG potential consists of a LJ potential and an added Gaussian minimum:

\begin{displaymath}
V_{\text{LJG}}(r) = \dfrac{1}{r^{12}} - \dfrac{2}{r^{6}} - \epsilon \exp{\left(-\dfrac{(r-r_0)^2}{2\sigma^2}\right)}.
\end{displaymath}

\begin{centering}
\begin{figure}
\includegraphics[width=\textwidth] {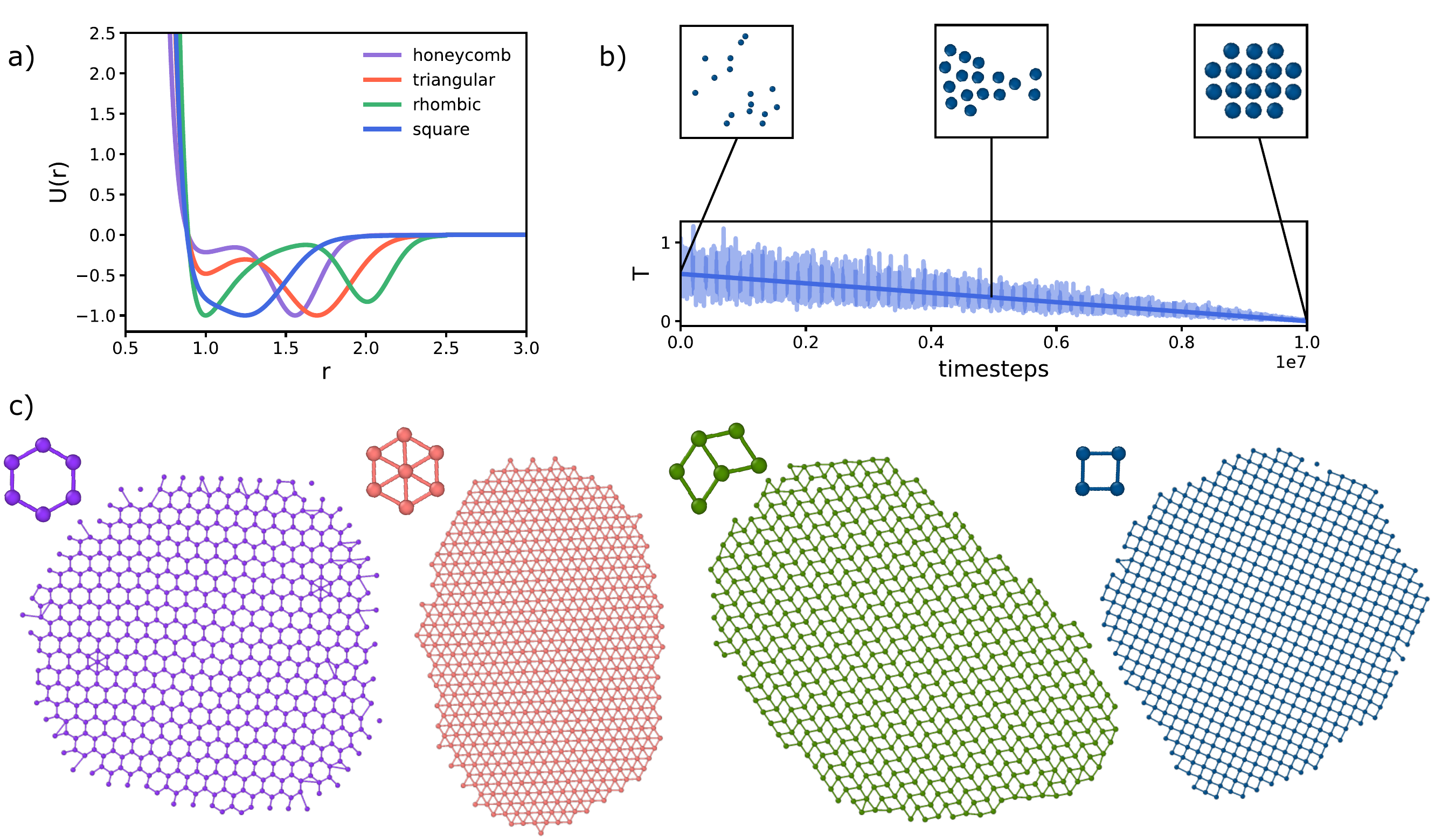}
\caption{
\textbf{State points and simulation methods.}
\textbf{(a)} LJG pair potentials for the four state points examined in this paper. 
The pair potentials result in the bulk self-assembly of the square, triangular, rhombic, and honeycomb lattice structures. 
\textbf{(b)} Quenching process inducing cluster formation, shown for the square lattice system at $N = 16$. 
The dark blue line is the set temperature and the light blue line is the measured temperature in the system.
The temperature is initially set at 0.6, then cooled to 0.01 in the $NVT$ ensemble over $10^7$ time steps. 
Cluster formation is illustrated via simulation snapshots at $0$, $0.5\times10^7$, and $10^7$ time steps, respectively. 
\textbf{(c)} Bulk crystals of the four systems, shown in clusters of size $N = 625$ (from left to right: honeycomb, triangular, rhombic and square lattices). 
For each bulk crystal, the corresponding tiling motif is shown in the top left corner.}
\label{fig1}
\end{figure}
\end{centering}

The expression above results in a double-well potential for particles separated by center-to-center distance $r$, where the first well is located at position 1 (in dimensionless units) and the second well has position $r_0$, depth $\epsilon$, and width $\sigma$.
We chose the variables  $r_0$, $\epsilon$, and $\sigma$ so as to study the formation of the square, triangular, honeycomb, and rhombic bulk lattice structures.\cite{Engel2007a,Engel2008a}
Table~\ref{tab:state_points} shows each chosen value of $r_0$, $\epsilon$, and $\sigma$, as well as the corresponding bulk structure, and its associated coordination number ($CN$).
The $CN$ of each bulk structure generally increases in pair potentials with deeper wells at the shorter length scale in two-well pair potentials.\cite{Pan2023}
The notable exception is the triangular lattice, which has the highest $CN=6$ despite a very shallow first well. %
In all cases, we normalize $V_\text{LJG}$ such that its minimum value is $-1$.
Fig.~\ref{fig1}a depicts the pair potentials that were investigated in this study.
Note that for this set of chosen pair potentials, the distance between the first and second potential wells increases monotonically as the potentials progress from forming square, to honeycomb, to triangular, to rhombic lattices, and the relative discrepancy in well depths (indicating asymmetry in bonding preference) decreases monotonically as the potentials progress from forming honeycomb, to triangular, to rhombic, to square lattices.
In the case of the square potential, the wells significantly overlap, resulting in one wide energy minimum with a shoulder.
The rhombic potential is the only case where the first well is deeper than the second well.

\renewcommand{\arraystretch}{1.3}
\begin{table}[h]
\caption{All state points used in this study, with corresponding bulk structures formed and their respective sp-$n$ rings and coordination numbers ($CN$s).}
\label{tab:state_points}
\centering
\begin{tabular}{|c|c|c|c|}
\hline
\textbf{State point} & \textbf{Bulk structure} & \textbf{Shortest-path ring} & \textbf{Coordination number} \\ \hline
$r_0=1.3$, $\epsilon =1.5$, $\sigma^2 =0.042$ & Square lattice & sp-4 & $CN=4$ \\ \hline
$r_0=1.7$, $\epsilon =2.0$, $\sigma^2 =0.042$ & Triangular lattice & sp-3 & $CN=6$\\ \hline
$r_0=1.56$, $\epsilon =4.5$, $\sigma^2 =0.02$ & Honeycomb lattice & sp-6 & $CN=3$ \\ \hline
$r_0=2.01$, $\epsilon =0.8$, $\sigma^2 =0.02$ & Rhombic lattice & sp-3 & $CN=6$\\
\hline

\end{tabular}
\end{table}

We also simulated the formation of finite clusters ranging from $N=2$ to $100$ particles for the simpler system governed by the single-well LJ potential:
\begin{displaymath}
V_{\text{LJ}}(r) = \dfrac{1}{r^{12}} - \dfrac{2}{r^{6}}.
\end{displaymath}
The system governed by $V_{\text{LJ}}$ only exhibits simple, close-packed cluster structures in two dimensions, with the one feature of the interaction potential defining the bond length.\cite{Nelson1989}
We used these clusters as references when examining cluster growth sequences for the multi-well LJG systems.
For numerical reasons, we note that we set $V_\text{LJG} = 0$ and $V_\text{LJ} = 0$ at all values of $r < 0.5$ and $r \geq 2.5$.

All molecular dynamics simulations were performed using the open-source simulation package \emph{HOOMD-blue}.\cite{Anderson2020} 
We managed simulation data using the \emph{signac} data management framework \cite{Adorf2018,Ramasubramani2018} and visualized all simulations with OVITO.\cite{Stukowski2010} 
Analysis was performed using the analysis packages \emph{freud} \cite{Ramasubramani2020a} and \emph{NetworkX}.\cite{Hagberg2008}

Cluster formation was promoted by fully quenching each system from the high temperature of $0.6\,kT$ (in dimensionless units) to the low temperature of $0.01\,kT$ over $10^7$ time steps (Fig.~\ref{fig1}b).
All simulations were performed in the isochoric--isothermal (canonical, or $NVT$) ensemble with time step $dt=0.005$ and Nos\'e--Hoover thermostat coupling constant $\tau=1$. 
Five replica simulations were generated for each system size.
In addition, we simulated 21--23 replicas of larger clusters of $N = 625$ for each system, to confirm the self-assembly structures of all state points and compare to larger system sizes approaching bulk ordering (Fig.~\ref{fig1}c).
For each quench, we chose the final frame of the simulation for further analysis.

\subsection{Analysis}

\subsubsection{Radial distribution function}

For all clusters, we calculated radial distribution functions (RDFs) to track the emergence of bulk-like structure as a function of system size. 
The RDF $g(r)$ is a statistical description of how particle density varies as a function of distance $r$ from a central particle.
It is calculated as:
\begin{displaymath}
    g(r) = V \frac{N-1}{N} \langle \delta(r) \rangle,
\end{displaymath}
where $V$ is system volume, $N$ is the number of particles, the ensemble average is taken over all particle pairs $i \neq j$ in the system, and $\delta(r)$ is the Kronecker delta.
$g(r)$ thus peaks at distances that characterize average nearest-neighbor shells around particles. 

\subsubsection{Shortest-path rings}

To analyze the emergence of specific motifs associated with bulk structure in each system, we tracked the formation of so-called ``shortest-path rings" in all clusters.\cite{Franzblau1991}
These rings are topological signatures of order and reflect information related to the underlying bond network of each cluster: within the bond network, shortest-path rings are cycles that have no shortcuts across them.
To identify these rings, we first construct a graph $G=(V,E)$ consisting of a set of $V$ vertices and $E$ edges associated with each cluster. 
Each particle in the cluster is a vertex and an edge is generated between particles $i$ and $j$ if the distance between them, $r_{ij}$, satisfies the criterion 
$r_{\text{min}} \leq r_{ij} \leq r_{\text{max}}$.
For all systems, we set the minimal bond distance $r_{\text{min}} = 0$, and the maximal bond distance $r_{\text{max}} = 1.3$, in order to capture bonding associated with the first well of each LJG potential.
Note that, since we only monitor motifs formed from bonds with distances $r \leq 1.3$, we do not capture structural motifs associated with longer bonds.

We find the shortest-path rings of size $n$ in each network by first identifying all chordless cycles with $n$ edges, then identifying the subset of chordless cycles $R$ for which $\text{dist}_R(i,j) = \text{dist}_G(i,j)$ for all pairs $(i,j)$ in each ring.
$\text{dist}_G(i,j)$ is the number of edges that makes up the shortest possible path between vertices $i$ and $j$ in graph $G$, and $\text{dist}_R(i,j)$ is the number of edges that makes up the shortest possible path between vertices $i$ and $j$ in cycle $R$.
In this paper, we focus on shortest-path rings with $n=(3,4,5,6)$ edges, abbreviated as sp-3, sp-4, sp-5, and sp-6 rings, respectively.
These rings are topological proxies of triangular, square or rhombic, pentagonal, and hexagonal bonding motifs in which only connectivity matters.
The bulk structures stabilized by the pair potentials of this paper are each characterized by one ring type (Table~\ref{tab:state_points}).
An advantage of using a topological measure such as rings to identify structural features, rather than other metrics that might rely on bond angles or strict distance cutoffs, is their robustness against slight alterations in local structure.
This flexibility was necessary since each cluster contains structural features that fluctuate significantly with system size, especially for small $N$.

\subsubsection{Bonding energy}

We also monitored changes in bonding as a function of system size by calculating total potential energy associated with bonding on multiple length scales in each cluster.
Within each cluster, we calculated
\begin{displaymath}
    E(r_{\text{min}},r_{\text{max}}) = \sum_{r_{\text{min}} \leq r_{ij} \leq r_{\text{max}}} V_\text{LJG}(r_{ij}),
\end{displaymath}
where the sum proceeds over each particle pair $(i,j)$ in the system whose center-to-center distance, $r_{ij}$, meets the criterion $r_{\text{min}} \leq r_{ij} \leq r_{\text{max}}$.
Each value of $E(r_{\text{min}},r_{\text{max}})$ thus represents the portion of the cluster's potential energy associated with bonds whose lengths are within the range $(r_{\text{min}}, r_{\text{max}})$.  
We examined two bonding regimes for clusters in all systems: $(r_{\text{min}}, r_{\text{max}}) = (0,1.3)$, and $(r_{\text{min}}, r_{\text{max}}) = (1.3,2.5)$.
The first regime captures the effects of bonding within the first energy well of $V_\text{LJG}$ for all systems, while the second regime captures the effects of longer-range bonding associated with the second energy well.
We also calculated the total potential energy $E(0,2.5) = E(0,1.3) + E(1.3,2.5)$ of each cluster.

To better understand the growth sequence of each system, we compared the energy per particle of each cluster, $E/N$, to the value of $E/N$ which would result from the minimal energy packing of $N$ non-overlapping disks of diameter $\sigma$ with the following pair potential:
\begin{displaymath}
    V_\text{cp}(r) = 
    \begin{cases}
    -1 & \text{if } r = \sigma\\
    0 & \text{if } r > \sigma.
    \end{cases}
\end{displaymath}
The potential $V_\text{cp}$ is ``sticky" and only rewards bonding at a single length scale, thus giving rise to close packings in two dimensions that minimize energy---also termed penny packings.\cite{Graham1990}
It is known that these penny packings are subsets of the triangular lattice; however, the precise structural sequence and energies of these clusters with changes in $N$ has been an area of active investigation.\cite{Graham1990,Chow1995,Bernstein1997}
In these studies, the second moment of each cluster is typically minimized rather than the total energy; these optimal clusters are still called ``minimum energy clusters" since they represent closest packings.\cite{Graham1990,Sloane1995}

We used the minimal second-moment cluster sequence determined in Ref.~\citenum{Graham1990} as a proxy for closest-packed clusters and calculated $E/N$ as a function of $N$ for this sequence.
We chose the bonding energy of $V_\text{cp}$ to be $-1$ in order to be consistent with the energy of the deepest well of $V_\text{LJG}$ and $V_\text{LJ}$, so that we could directly compare energies of bonding at the length scale associated with the deepest well to energies of bonding if the system exclusively packed closely at that length scale.

\section{Results}

\subsection{The Lennard-Jones system}
\begin{centering}
\begin{figure}
\includegraphics[width=\textwidth] {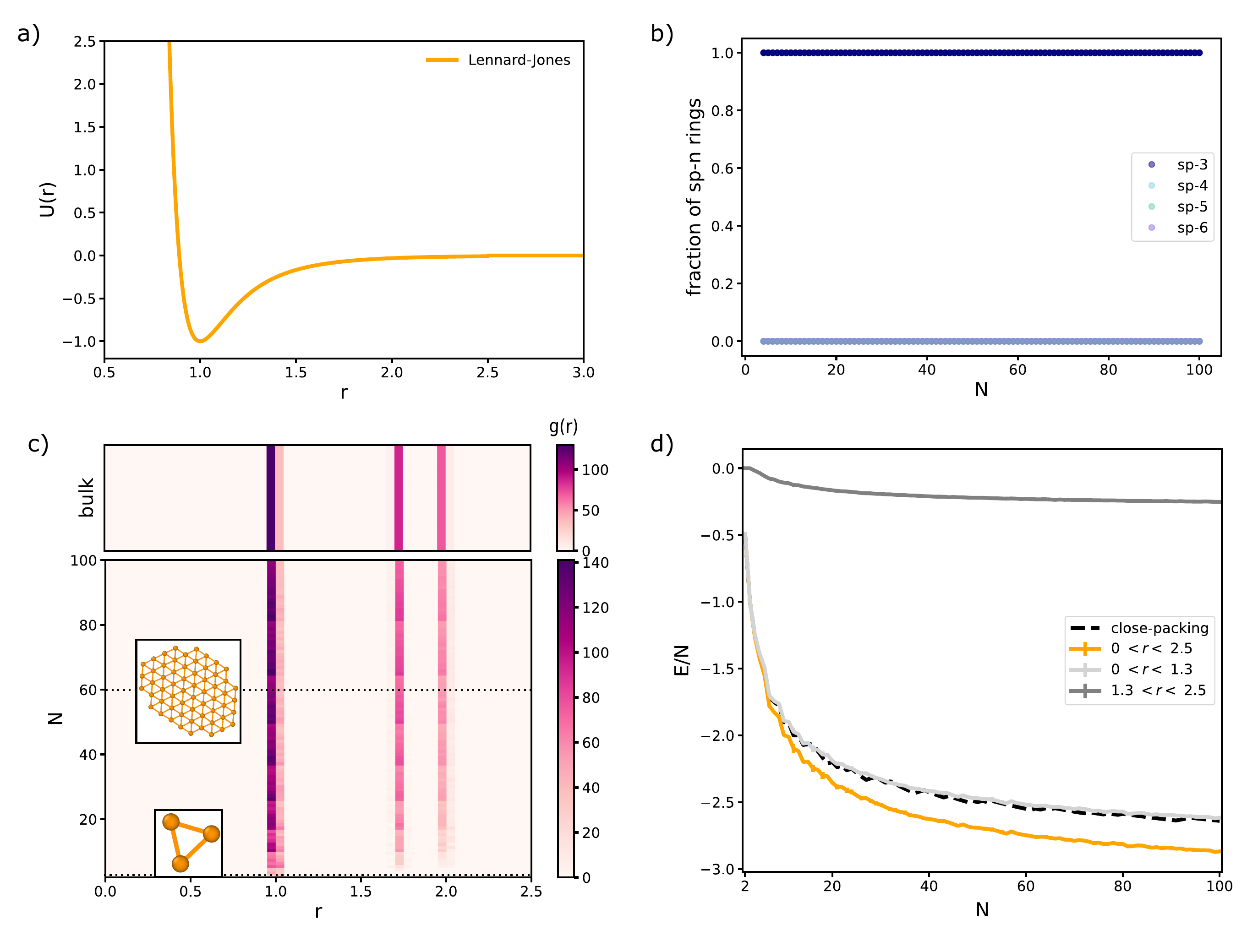}
\caption{
\textbf{In the 2D LJ system, the bulk structure appears immediately, even within clusters of small size $N$.}
\textbf{(a)} The LJ pair potential. 
\textbf{(b)} sp-3, sp-4, sp-5, and sp-6 ring fractions in each LJ cluster as a function of $N$. 
sp-4, sp-5, and sp-6 ring fractions all overlap at the constant value 0 for all $N$.
Within each cluster, ring fractions are computed with respect to the total number of all sp-3, sp-4, sp-5, and sp-6 rings in that cluster.
At each value of $N$, ring fractions of five replicate clusters are shown as separate (but overlapping) data points.
\textbf{(c)} Heat map showing the radial distribution function $g(r)$ for each LJ cluster as a function of $N$.
At each value of $N$, $g(r)$ is averaged over five replicate clusters.
The average $g(r)$ taken over 23 replicate clusters with system size $N=625$, representing the bulk structure, is also shown as a separate panel at the top.
Insets show minimal-energy clusters at $N = 3$ and $N=60$.
\textbf{(d)} Potential energy per particle in each cluster, $E/N$, as a function of $N$. 
Values of $E/N$ are shown for three bonding regimes: nearest-neighbor bonds with distance  $0 \leq r \leq 1.3$, next-nearest neighbor bonds with distance $1.3 \leq r \leq 2.5$, and all bonds (with distance $0 \leq r \leq 2.5$).
Each value shown is an average taken over five replicate clusters, and error bars indicate the standard deviation across replicas. 
A dotted line also indicates $E/N$ for perfectly close-packing.}
\label{LJfig2}
\end{figure}
\end{centering}

We first examine the emergence of bulk structure in finite clusters of particles interacting via the LJ potential $V_\text{LJ}$ (Fig.~\ref{LJfig2}a), which provides a  benchmark against which we will compare cluster formation in the more complicated LJG systems. 
The LJ system produces simple, close-packed minimum energy clusters commensurate with the bulk structure at all system sizes. 
We verify the generation of these closed-packed structures at all values of $N$ through analyses of shortest-path rings, $g(r)$, and the potential energy of each cluster.

At all values of $N \geq 3$, the fraction of sp-3 rings in each cluster is 1, 
indicating that each cluster's bond network exclusively contains sp-3 rings---topological proxies for triangular motifs (Fig.~\ref{LJfig2}b). 
The radial distribution function $g(r)$ shows a well-defined first peak at $r = 1$ at the lowest system size, and this peak remains at $r = 1$ at all values of $N$, showcasing uniformity in cluster structure as system size increases (Fig.~\ref{LJfig2}c). 
Other peaks corresponding to next-nearest neighbor distances emerge at larger system sizes and similarly remain constant with increasing $N$.
The potential energy per particle, $E/N$, of each cluster directly verifies that the LJ system follows a close-packing growth sequence
(Fig.~\ref{LJfig2}d).
Nearest-neighbor bonding $E(0,1.3)/N$ almost exactly follows $E/N$ for perfectly close-packing, indicating that, as $N$ increases, the system forms minimal energy clusters associated with bonding at a single length scale.
Note that the potential energy of the entire system, $E(0,2.5)/N$, becomes lower than $E/N$ for perfectly close-packing at larger values of $N$.
This is due to the relatively large width of the energy well in $V_\text{LJ}$, which means that next-nearest neighbor bonds contribute to the potential energy of each cluster as $N$ increases. 
The emergence of these bonds can be seen in the plot of $E(1.3,2.5)/N$ as a function of $N$. 

\subsection{The Lennard-Jones--Gauss system: Bond lengths}

\begin{centering}
\begin{figure}
\includegraphics[width=\textwidth] {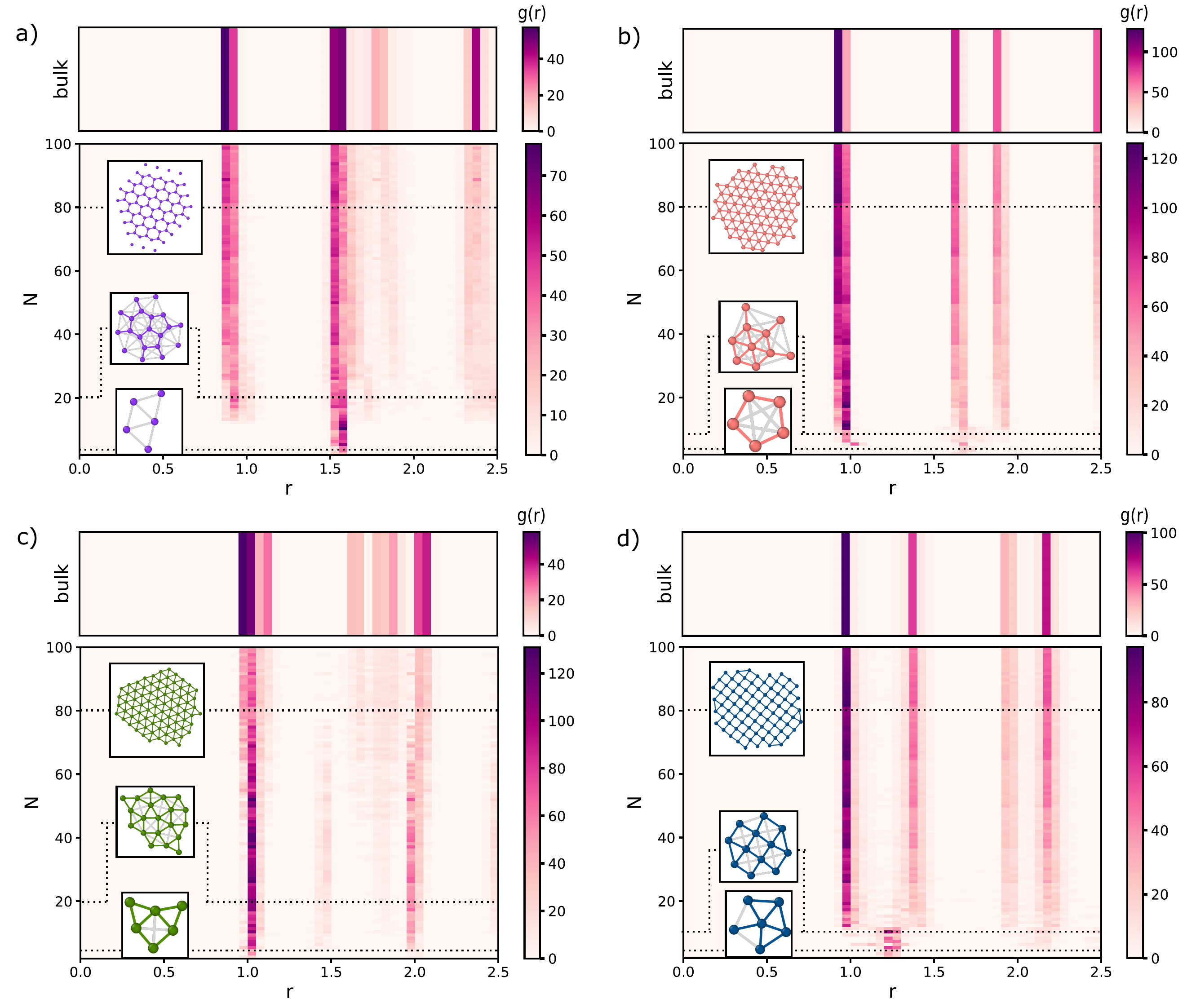}
\caption{
\textbf{Radial distribution functions show bonding lengths associated with cluster geometries.}
Each panel corresponds to a different system and shows a heat map of $g(r)$ as a function of system size $N$, $g(r)$ for the corresponding bulk structure at $N=625$, and snapshot insets of example clusters at varying values of $N$. 
For $2<N<100$, values of $g(r)$ are averaged over five replicas at each system size, and for $N=625$, values of $g(r)$ are averaged over 21 replicas.
\textbf{(a)} The honeycomb system, with inset snapshots of cluster structures at $N=5, 20, 80$. 
\textbf{(b)} The triangular system, with inset snapshots of cluster structures at $N=5, 10, 80$.  
\textbf{(c)} The rhombic system, with inset snapshots of cluster structures at $N=6, 20, 80$.  
\textbf{(d)} The square system, with inset snapshots of cluster structures at $N=6, 12, 80$.
In all cases, bonds are colored according to system and bond length:
all bonds with length $1.3<r<1.9$ are colored gray, and bonds with length $r<1.3$ are colored according to the system.
(At the largest system sizes, bonds of length $1.3<r<1.9$ are not shown for clarity.)
}
\label{rdf}
\end{figure}
\end{centering}

In contrast to the LJ system, LJG systems with two competing preferred length scales show emergence of structure with more complicated bonding motifs as system size increases. 
The lengths of the bonds which form these motifs can be seen in heat maps of $g(r)$ for all clusters of $N = 2$ to $100$ particles (Fig.~\ref{rdf}).
In all systems, $g(r)$ shows fluctuations at smaller system sizes before stabilizing at larger system sizes.
Distributions of $g(r)$ at larger system sizes, above $N=80$, match the distribution of $g(r)$ for the bulk $N=625$ in all cases. 
Bonds of different lengths emerge in each system at values of $N$ which depend upon the shape of the pair potential, with deeper potential wells causing bonds associated with those wells to emerge at smaller system sizes.

The deepest wells of the honeycomb, triangular, and rhombic potentials are located at $r=1.5, 1.7$, and 1, respectively.
Accordingly, bonding at these length scales emerges immediately in each system at the smallest values of $N$.
Bonding at the additional length scale associated with the shallower potential well depends on the depth of that well: greater depth causes associated bonding to occur at smaller $N$.
For the honeycomb potential (with the shallowest well, at $r=1$), bonding at this distance emerges at $N = 12$ particles; for the triangular potential (with a slightly deeper well, also at $r=1$), bonding at this length scale occurs at $N = 4$ particles; and for the rhombic potential (with a deeper well still at $r=2$), bonding at $r=2$ occurs at $N = 4$.

The square potential consists of a shoulder at $r=1$ and minimum at $r=1.3$, rather than two well-separated minima, and therefore displays unique bond length behavior with increasing $N$.
At small system sizes, bonding at $r=1.3$, associated with the minimum of the potential, is preferred.
Bonding distances then `split' once the system size surpasses $N=12$, resulting in bonding at $r=1$ and $r=1.4$.
Due to the large width of the square potential's well, bonding at the non-optimal distances (\textit{i.e.}, not at the minimum at $r=1.3$) are accommodated once the system is sufficiently large.

\subsection{The Lennard-Jones--Gauss system: Bond topology}

\begin{centering}
\begin{figure}
\includegraphics[width=\textwidth] {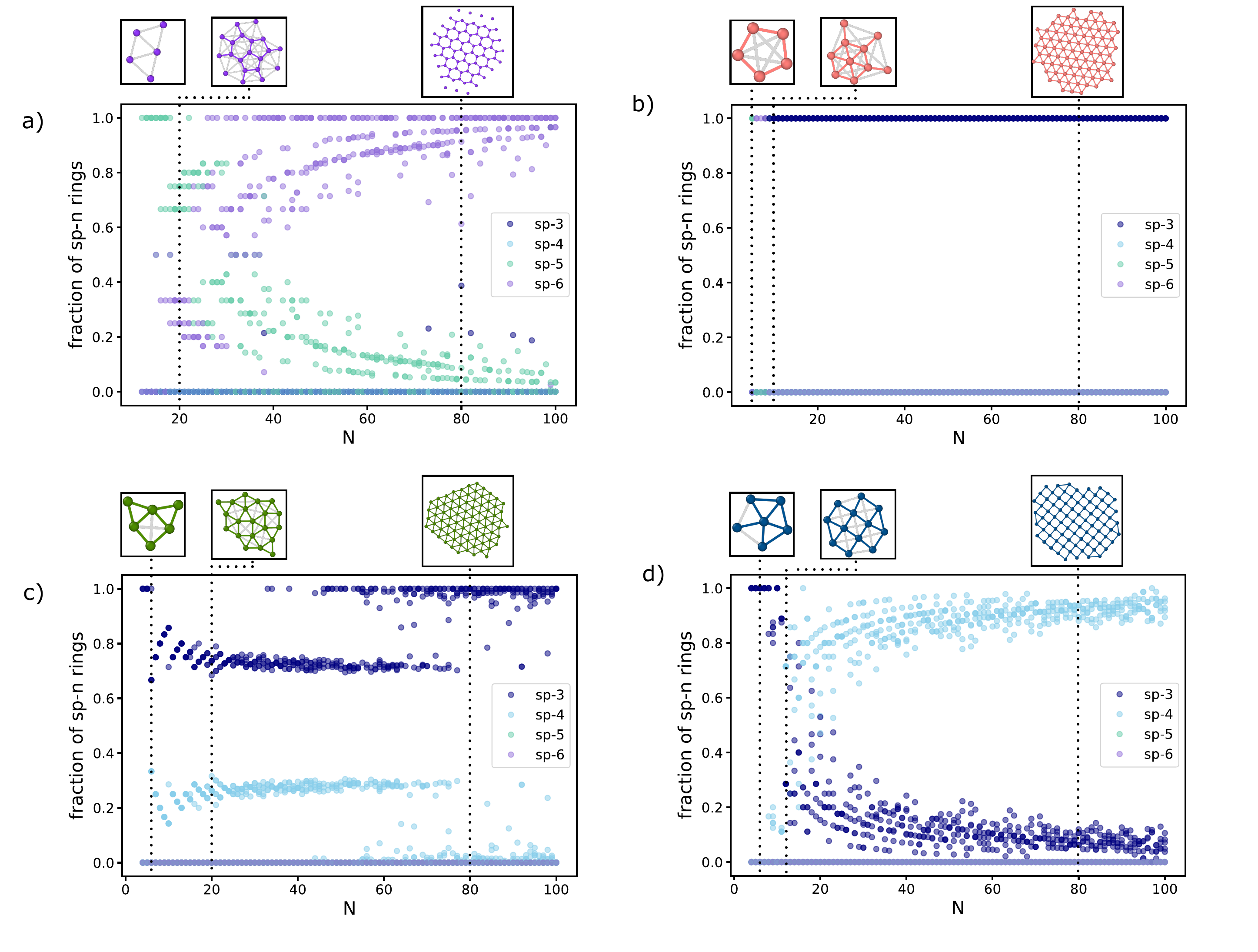}
\caption{ 
\textbf{Shortest-path rings show bonding topology associated with cluster geometries.}
Each panel corresponds to a different system and shows the fraction of rings that are sp-3, sp-4, sp-5, and sp-6 as a function of system size $N$.
Fractions are computed with respect to the total number of all sp-3, sp-4, sp-5, and sp-6 rings in each cluster.
At each value of $N$, ring fractions of five replicate clusters are shown as separate data points. 
Panels also show example cluster snapshots for each system.  
\textbf{(a)} The honeycomb system, with inset snapshots of cluster structures at $N=5, 20, 80$. 
\textbf{(b)} The triangular system, with inset snapshots of cluster structures at $N=5, 10, 80$. 
\textbf{(c)} The rhombic system, with inset snapshots of cluster structures at $N=6, 20, 80$. 
\textbf{(d)} The square system, with inset snapshots of cluster structures at $N=6, 12, 80$. 
In all cases, bonds are colored according to system and bond length: all bonds with length $1.3<r<1.9$ are colored gray, and bonds with length $r<1.3$ are colored according to the system.
(At the largest system sizes, bonds of length $1.3<r<1.9$ are not shown for clarity.)
}
\label{ring}
\end{figure}
\end{centering}

To gain further insight into emergent cluster structures, we examine the bonding topology in each system by tracking the shortest-path (sp) rings in clusters as a function of system size, as shown in Fig.~\ref{ring}.
In each system, the bulk structure's ring motif eventually dominates at large system sizes.
For the honeycomb, triangular, rhombic, and square systems, the ring motifs associated with each bulk structure are the sp-6 (hexagonal) ring, sp-3 (triangular) ring, sp-3 (triangular) ring, and sp-4 (square) ring, respectively (Tab.~\ref{tab:state_points}). 

The relative proportions of ring motifs at intermediate system sizes provide insight into how bulk structure emerges with increasing $N$.
The honeycomb and square systems both display a gradual switch in sp-$n$ ring propensity as they shift towards bulk structure. 
In the honeycomb system, there is a shift from sp-5 to sp-6 ring dominance as $N$ increases, indicating a transition from mostly pentagonal to mostly hexagonal ring motifs between $20<N<40$.
In the square system, there is a shift from sp-3 to sp-4 ring dominance, indicating a transition from mostly triangular to mostly square ring motifs, before $N=20$. 
The transition to the bulk motif occurs at even smaller $N$ in the triangular system: for $N<10$ there are some sp-3, sp-5, and sp-6 rings, but these give way to only sp-3 rings for $N >10$. 

The rhombic system shows a markedly different pattern of structural emergence as a result of competing structures at intermediate cluster size.
At very small $N$ rings are exclusively sp-3, with sp-4 rings emerging at $N=6$.
Thereafter, until around $N=80$, a distinct structure defined by a minority of sp-4 rings and a majority of sp-3 rings dominates.
These clusters are subsets of the Archimedean snub square tiling, comprised of ordered squares and triangles. 
The rhombic bulk structure emerges around $N=50$, and becomes dominant for $N > 80$: this structure is characterized entirely by sp-3 rings, since it is a rhombic tiling of staggered triangles.
We performed additional analyses of bond angle distributions as a function of $N$ to further investigate structure emergence in each of our systems; details and an extended discussion of those results can be found in the \emph{Supplementary Information}.

\subsection{The Lennard-Jones--Gauss system: Cluster energies}

\begin{centering}
\begin{figure}
\includegraphics[width=\textwidth]{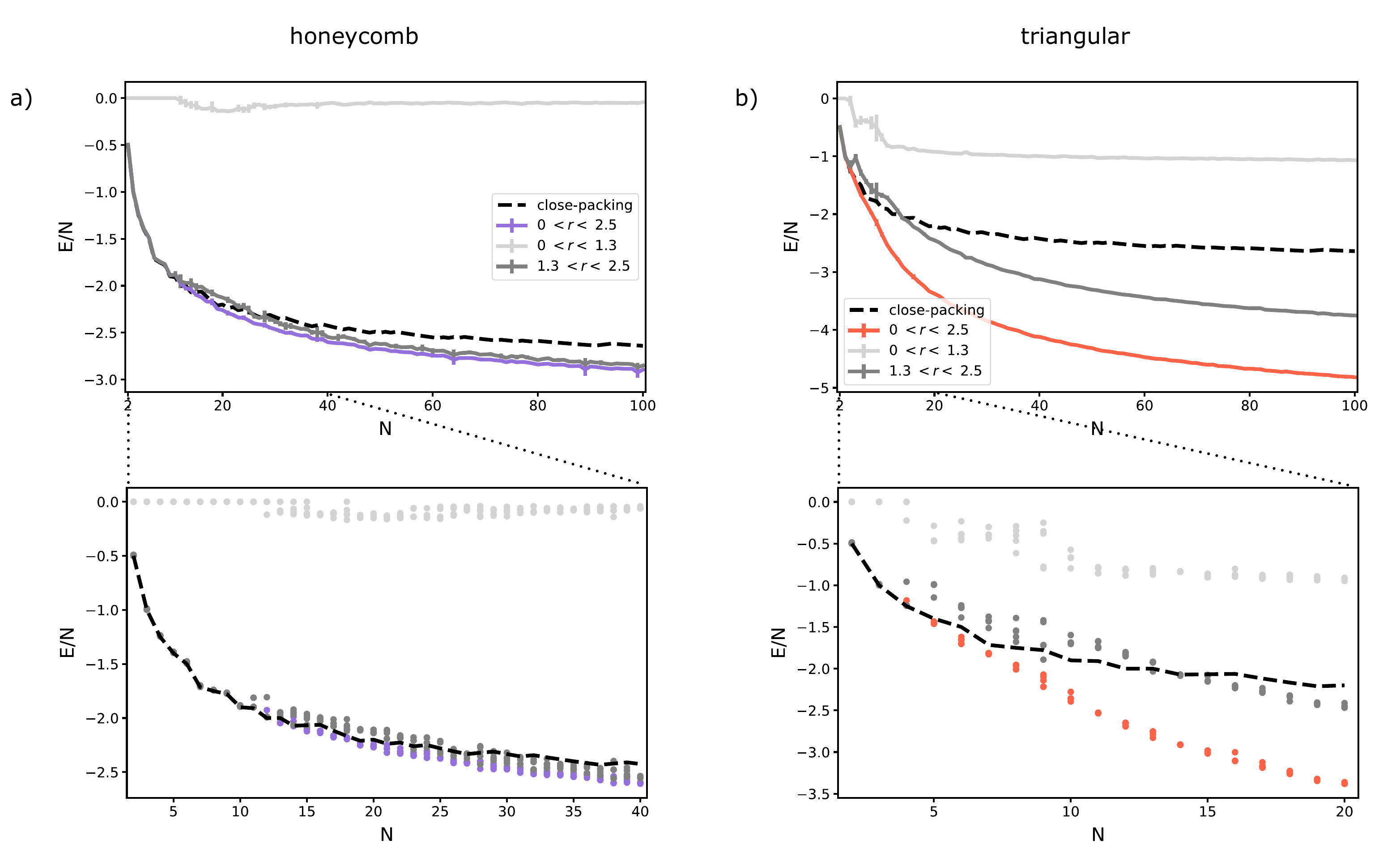}
\caption{
\textbf{Potential energy per particle of clusters in the honeycomb and triangular systems.}
\textbf{(a)} Potential energy per particle, $E/N$, as a function of system size $N$, for the honeycomb system.
\textbf{(b)} Potential energy per particle, $E/N$, as a function of system size $N$, for the triangular system.
Different colors correspond to energies associated with different bond lengths: light gray with shorter bonds, dark gray with longer bonds, and color (according to the system) with all bonds. 
The $E/N$ sequence that corresponds to perfectly close-packing is also shown as a black dotted line, for purposes of comparison.
In all panels, top plots show the full range of $N$ considered in this study and contain $E/N$ averaged over five replicas at each system size. 
Error bars show the standard deviation across replicas.
Bottom plots show zoom-ins on $N$ ranges of interest, with replicas as individual data points.
}
\label{energy1}
\end{figure}
\end{centering}

\begin{centering}
\begin{figure}
\includegraphics[width=\textwidth]{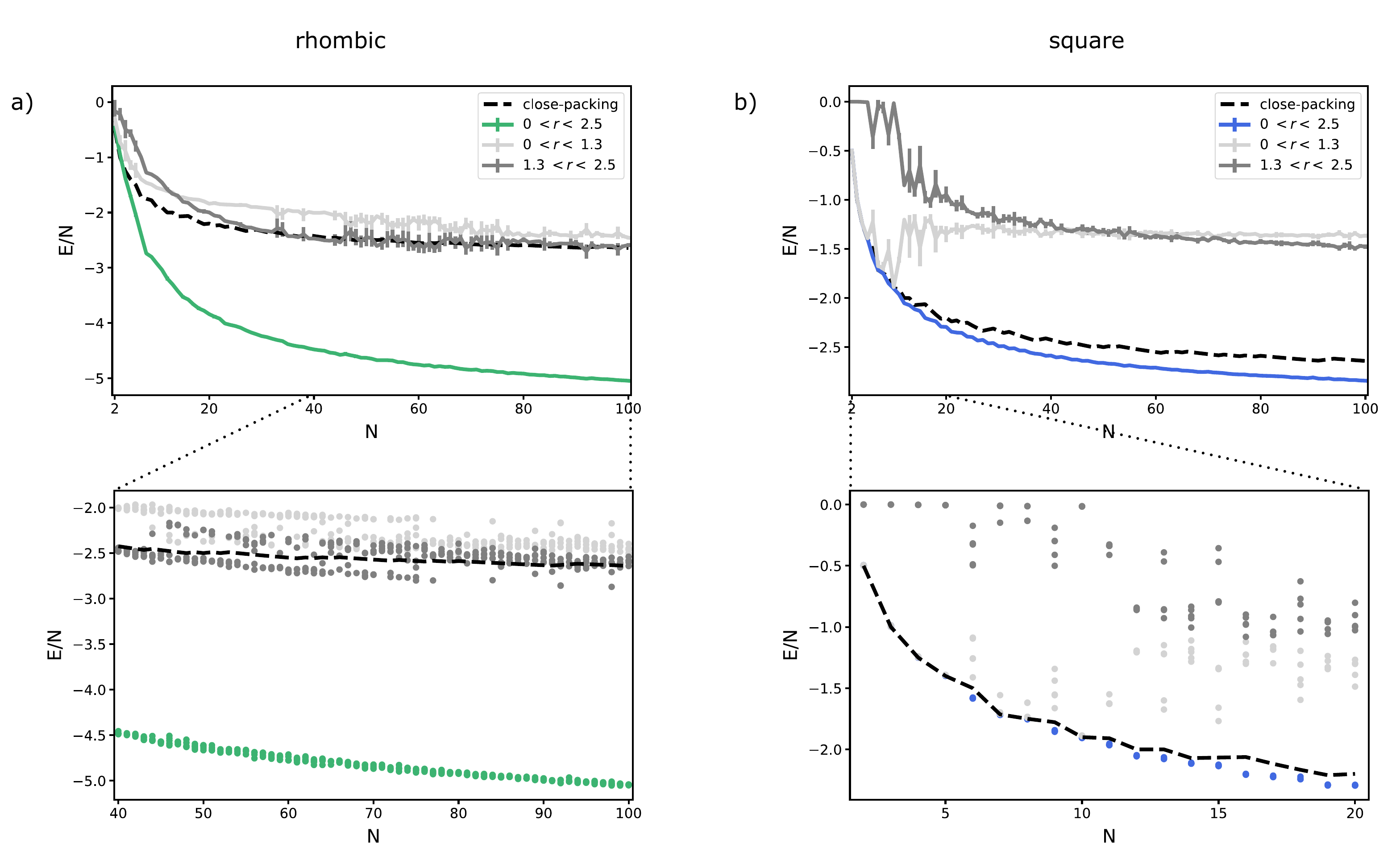}
\caption{
\textbf{Potential energy per particle of  clusters in the rhombic and square systems.}
\textbf{(a)} Potential energy per particle, $E/N$, as a function of system size $N$, for the rhombic system.
\textbf{(b)} Potential energy per particle, $E/N$, as a function of system size $N$, for the square system.
Different colors correspond to energies associated with different bond lengths: light gray with shorter bonds, dark gray with longer bonds, and color (according to the system) with all bonds. 
The $E/N$ sequence that corresponds to perfectly close-packing is also shown as a black dotted line, for purposes of comparison.
In all panels, top plots show the full range of $N$ considered in this study and contain $E/N$ averaged over five replicas at each system size. 
Error bars show the standard deviation across replicas.
Bottom plots show zoom-ins on $N$ ranges of interest, with replicas as individual data points.
}
\label{energy2}
\end{figure}
\end{centering}

Finally, we investigate structure emergence from an energetic perspective.
For each cluster size, we examine potential energy $E(r_{\text{min}},r_{\text{max}})$ associated with bond lengths $r_{\text{min}} < r < r_{\text{max}}$ for three bonding regimes: short bonds $(0,1.3)$, long bonds $(1.3,2.5)$, and all bonds $(0,2.5)$.
Figs.~\ref{energy1} and \ref{energy2} show potential energy per particle, $E/N$, as a function of $N$ and compare to the equivalent $E/N$ for perfectly close-packing.

For all systems, $E(0,1.3)$ plateaus at smaller system sizes than $E(1.3,2.5)$. 
This is because the energy contribution from shorter bonds saturates at small system sizes, while the energy contribution from longer bonds continues to grow as system size increases and saturates at higher $N$.
The plateau in $E/N$ from long bonds generally corresponds to the formation of the bulk structure: at $N\approx 30$ in the honeycomb system (Fig.~\ref{energy1}a), $N\approx 10$ in the triangular system (Fig.~\ref{energy1}b), $N\approx 80$ in the rhombic system (Fig.~\ref{energy2}a), and $N\approx 12$ in the square system (Fig.~\ref{energy2}b).
The rhombic system is especially noteworthy: a zoom-in of $E/N$ at larger system sizes shows a clear transition from the snub square tiling to the rhombic tiling (Fig.~\ref{energy2}a, lower panel).
The snub square tiling regime consists of separate bands for $E(0,1.3)/N$ and $E(1.3,2.5)/N$ as a function of $N$, with the longer bonds associated with lower potential energies.
By contrast, the rhombic structure which emerges at $N=46$ is  characterized by similarity in the $E(0,1.3)/N$ and $E(1.3,2.5)/N$ bands as a function of $N$. 

At low values of $N$, all systems exhibit close-packing at the length scale associated with the deepest well of the governing pair potential, indicated by the correspondence between the measured $E/N$ and $E/N$ for perfect close-packing as a function of $N$.
In both the honeycomb and triangular systems, $E(1.3,2.5)/N$ follows close-packing energy minimization until bonding at the shorter distance $0< r < 1.3$ appears (Fig.~\ref{energy1}). 
For the honeycomb system, this transition occurs at $N=12$, while for the triangular system it occurs at $N=4$.
When bonding at the shorter length scale first appears at these system sizes, longer bonds at $1.3 < r < 2.5$ become more energetically expensive than close-packing, but clusters have lower total energy for all interparticle distances $0 < r < 2.5$ compared to the energy associated with close-packing.

In both the rhombic and square systems, $E(0,1.3)/N$ follows close-packing for $N=2$--$3$ and $N=2$--$5$, respectively (Fig.~\ref{energy2}).
Bonding at the longer length scale $1.3< r < 2.5$ is introduced at very low $N$ in these systems, causing bonding at the shorter length scale to become more energetically expensive than close-packing, but also rendering a lower total energy for all interparticle distances $0 < r < 2.5$ compared to the close-packing energy.
For the rhombic system at high values of $N$, the value of $E(0,1.3)/N$ does again approach that of the close-packing, which is sensible given that the rhombic structure is essentially a staggered triangular tiling.  

\section{Discussion}

Taken together, our analyses uncover a portrait of structure emergence within finite clusters that is dependent on governing pair potential shape and the geometry of the bulk structure.
First, we describe the structure emergence in each system, showcased in the example cluster snapshots in Figs.~\ref{rdf} and \ref{ring}.
In the honeycomb system, initial bonding around $r=1.5$ results in closest-packed subsets of the triangular lattice.
Pentagonal (sp-5) ring motifs emerge when bonding at the shorter length $r = 1$ appears at $N = 12$, and these eventually morph to hexagonal (sp-6) ring motifs at larger $N$ (when the peak in $g(r)$ around $r=1.8$ appears).
In the triangular system, there is some closest-packing at low $N$ associated with bonding around $r=1.7$, but as soon as bonding around $r=1$ emerges, the clusters display a mix of open and closed structures associated with bonding at both length scales and mixed sp-$n$ rings.
Subsets of the triangular lattice do not appear until $N=8$ (when sp-3 rings dominate and the peak in $g(r)$ around $r=1.9$ appears). 
In the rhombic system, bonding at $r=1$ and $r=2$ is present even at the smallest system sizes.
Square (sp-4) and triangular (sp-3) rings appear at $N=6$ and form subsets of the Archimedean snub square tiling.
Evidence of this tiling can also be found in the peak of $g(r)$ at around $r=\sqrt{2} \approx 1.41$, corresponding to the diagonal distance across the square. 
At around $N=50$, the rhombic lattice (characterized exclusively by sp-3 rings and a new peak of $g(r)$ around $r=1.7$) emerges, and becomes dominant for $N>80$.
Finally, in the square system, bonding at $r=1.3$ for small system sizes begets triangular (sp-3) ring motifs, and square (sp-4) ring motifs emerge consistently around $N=12$, when the system shows bonding at $r=1$ (and $r=\sqrt{2}$).

System sizes for which cluster structures transition to a subset of the bulk structure vary, with the more open structures---those with lowest number of bonds counted by the coordination number ($CN$) as seen in Tab.~\ref{tab:state_points}---transitioning at larger system sizes.
Within the honeycomb system ($CN=3$), this transition happens around $N=30$; within the square system ($CN=4$), this transition happens around $N=12$, and within the triangular system ($CN=6$), this transition happens around $N=10$.
The rhombic system (also $CN=6$) presents a special case, due to the apparent competition between two structure types at intermediate system sizes: in this system, there is first a transition to the more open snub square tiling ($CN=5$) around $N=6$, and then a transition at much higher system sizes ($N \approx 50$--$80$) to the more closed rhombic structure.

In all systems prior to the transition to the bulk structure, clusters follow close-packing at the (primary) length scale associated with the deepest well in their governing pair potential, before bonding at the other (secondary) length scale is introduced at larger $N$.
Introduction of bonding at a secondary length scale causes bonding at the primary length scale to be less energetically efficient than close-packing. 
Universally, the introduction of bonding at a secondary length scale also causes the system as a whole to be more energetically efficient than close-packing at a single length scale.
The value of $N$ at which bonds at the secondary length scale appear is dependent on the depth of the associated well, with shallower secondary wells causing bonds to appear at higher values of $N$.
For the honeycomb system (with the shallowest secondary well), bonding at the secondary length scale occurs at $N=12$; for the triangular system (with the next most shallow secondary well), bonding at the secondary length scale occurs at $N=4$; and for the rhombic system (with the least shallow secondary well), bonding at the secondary length scale already occurs at $N=2$ in some replicas.
The square system presents a special case because its governing potential does not have two separate wells, but rather one wide well with a shoulder and a minimum.
As a result, bonding for the square system is generally more flexible.
Bonds associated with the potential minimum at $r=1.3$ form at low $N$, but these bonds are replaced at higher $N \geq 12$ by two types of non-optimal bonds: bonds associated with the potential shoulder at $r=1$ and bonds of length $r=\sqrt{2}$.

\section{Conclusions}

Our results show that tuning interaction potentials can design for growth sequences within clusters on the way to the bulk structure, in addition to designing the bulk crystals themselves.
Through manipulating the depths and positions of two competing length scales, we were able to alter the system size at which bonding at each length scale---as well as the bulk structure---emerges.
As an interesting corollary result in our study, we found that the bulk structures in two of our systems---honeycomb and rhombic---are of higher energy than competing crystalline polymorphs (triangular and snub-square tiling) that occur at much lower frequency than the predominant bulk structure.
The \emph{Supplementary Information} contains an extended discussion of our observations.

Future work extending our analysis to three dimensions, where polytetrahedral frustration plays a role in small cluster formation, may also be illuminating: how would our observations change as a result of this additional frustration?
An examination of crystal growth \textit{via} particle-by-particle attachment would also serve as a complementary study: to what extent are the cluster structures observed here the result of cooperative ordering of many particles simultaneously, and could our results be reproduced by single-particle addition? 
Answers to these questions could provide insight into whether classical nucleation---which posits that particles attach individually onto growing crystallites---results in minimal-energy clusters at small system sizes.

Future directions enabled by this work can be applied to engineer soft matter clusters on the nano- or mesoscale for which bulk structure emerges immediately at small system sizes, or for which designed structural transitions occur at intermediate system sizes. 
Engineered clusters may thus be especially robust against environmental perturbation or possess structures that are especially sensitive and size-dependent, relevant for applications ranging from drug delivery to hierarchical materials design.

\section{Citation Diversity Statement}

Recent work has identified a bias in citation practices such that papers from women and other marginalized scholars in STEM are under-cited relative to expected rates. 
Here we sought to proactively consider choosing references that reflect the diversity of the field in thought, form of contribution, gender, and other factors. 
We use databases that store the probability of a name being carried by people of different genders to mitigate our own citation bias at the intersection of name and identity. 
By this measure (and excluding self-citations to the first and last authors of our current paper, and papers whose authors' first names could not be determined), our references contain 8.33\% woman(first)/woman(last), 14.58\% man/woman, 4.17\% woman/man, and 72.92\% man/man categorization. 
This method is limited in that names, pronouns, and social media profiles used to construct the databases may not, in every case, be indicative of gender identity. 
Furthermore, probabilistic studies of names cannot be used to detect citation costs that are specific to intersex, non-binary, or transgender people who are out to a large number of their colleagues. 
We look forward to future work that could help us to better understand how to support equitable practices in science.

\section{Acknowledgements}

This material is based upon work supported by the National Science Foundation under Grant No.\ DMR-2144094 and was supported in part by the Cornell Center for Materials Research with funding from the Research Experience for Undergraduates program (DMR-1757420), 
as well as the Camille and Henry Dreyfus Foundation through a Machine Learning in the Chemical Sciences and Engineering Award (ML-22-038). 
J.\ E.\ D.\ and E.\ G.\ T.\ acknowledge support from the Claudine Malone `63 Summer Science Research Scholars Gift, and J.\ E.\ D.\ acknowledges additional support from the Levitt Fellowship. 
M.\ M.\ M.\ acknowledges support from the National Science Foundation Graduate Research Fellowship Grant No.\ DGE-2139899 (2021--2024) and the Dolores Zohrab Liebmann Fellowship.

\section{References}
\bibliographystyle{apsrev4-2}
%\bibliography{submitted_paper}
\input{cluster_assembly.bbl}

\end{document}